\documentclass[prl,twocolumn,superscriptaddress,showpacs]{revtex4}

\usepackage[dvips]{graphicx}

\renewcommand{\vec}[1]{\mathbf{#1}}

\newcommand{\eref}[1]{(\ref{#1})}

\renewcommand{\(}{\left(}
\renewcommand{\)}{\right)}

\begin{document}
\title{Role of cross-helicity in magnetohydrodynamic turbulence} 
\author{Jean Carlos Perez}
\author{Stanislav Boldyrev}
\affiliation{Department of Physics, University of Wisconsin-Madison,
  1150 University Ave, Madison, WI 53706, USA}  
\date{\today}

\begin{abstract}
Strong incompressible three-dimensional magnetohydrodynamic turbulence
is investigated by means of high resolution direct numerical
simulations. The simulations show that the configuration space is
characterized by regions of positive and negative cross-helicity,
corresponding to highly aligned or anti-aligned velocity and magnetic
field fluctuations, even when the average cross-helicity is zero. To
elucidate the role of cross-helicity, the spectra and structure of
turbulence are obtained in `imbalanced' regions where cross-helicity
is non-zero. When averaged over regions of positive and negative
cross-helicity, the result is consistent with the simulations of
balanced turbulence. An analytical explanation for the obtained
results is proposed.
\pacs{52.35.Ra}
\end{abstract}

\maketitle

\emph{Introduction.}---Magnetohydrodynamic (MHD) turbulence has been a
starting point for modeling large-scale turbulent motion of plasmas in
a wide variety of settings, ranging from laboratory experiments to
astrophysical systems, ~\cite[e.g.,][]{biskamp}.  When written in
terms of the Els\"asser variables the incompressible MHD equations
read
\begin{equation}
  \(\frac{\partial}{\partial t}\mp\vec v_A\cdot\nabla\)\vec
  z^\pm+\left(\vec z^\mp\cdot\nabla\right)\vec z^\pm = -\nabla P,
  \label{mhd-elsasser}
\end{equation}
where the Els\"asser variables are defined as $\vec z^\pm=\vec
v\pm\vec b$, $\vec v$ is the fluctuating plasma velocity, $\vec b$ is
the fluctuating magnetic field normalized by $\sqrt{4 \pi \rho_0}$,
${\bf v}_A={\bf B}_0/\sqrt{4\pi \rho_0}$ is the contribution of the
uniform magnetic field ${\bf B}_0$, $P=(p/\rho_0+b^2/2)$ includes the
plasma pressure $p$ and the magnetic pressure, and $\rho_0$ is the
background plasma density that we assume constant. It follows from
these equations that for $\vec z^\mp(\vec x,t)=0$, an arbitrary
function $\vec z^\pm(\vec x,t)=F^\pm(\vec x\pm\vec v_At)$ is an exact
nonlinear solution that represents a non-dispersive wave propagating
along the direction $\mp\vec v_A$. 
Nonlinear interactions are thus the result
of collisions between counter-propagating Alfv\'en waves.

The first phenomenological picture of MHD turbulence was proposed
independently by Iroshnikov~\cite{iroshnikov63,iroshnikov64} and
Kraichnan~\cite{kraichnan} (IK), who predicted an inertial range
scaling of the isotropic energy spectrum $E(k)\sim k^{-3/2}$. In this
picture, the spectral energy transfer at a given scale $\lambda\sim
1/k$ results from the cumulative effect of multiple weak interactions
between counter-propagating Alfv\'en wave packets moving along the
magnetic field of the large-scale energy containing eddies.  One
shortcoming of this phenomenology is that it is based on the
assumption of an isotropic spectral transfer, in clear contradiction
with recent results that reveal the anisotropic character of MHD
turbulence~\cite[e.g.,][]{Shebalin,Montgomery81,biskamp}.  Indeed,
Galtier et.al.~\cite{galtier} applied the formalism of weak turbulence
to equations \eref{mhd-elsasser}, demonstrating that the spectral
transfer is much more efficient in the field-perpendicular plane, and
derived a steeper energy spectrum $E(k_\perp)\propto k_\perp^{-2}$~,
where $k_\perp$ is the field perpendicular wave-number. This scaling
was originally predicted in
\citep{ng-bhattacharjee,bhattacharjee-ng,ng} based on more
phenomenological grounds.

As the cascade proceeds to smaller scales, the eddies become
progressively more elongated in the field-parallel direction, and the
nonlinear interaction becomes stronger. Eventually, the so called
`critical balance' condition of Goldreich and Sridhar
(GS)~\citep{GS95} is established. This condition states that the
turbulence is considered strong when there is a formal balance between
the crossing time of two interacting Alfv\'en wave packets and the
characteristic nonlinear interaction time, i.e., $k_{\|}B_0\sim
k_{\perp} b_{\lambda}$, where $k_{\|}$ is the typical field-parallel
wave-number of the fluctuations spectrum, and $b_{\lambda}$ is the rms 
magnitude of the fluctuations at the scale~$\lambda\sim
1/k_{\perp}$. The resulting scaling in the GS picture is
$E(k_\perp)\sim k_\perp^{-5/3}$.

The explosive growth of massively parallel computers in recent years has
made direct numerical simulations of MHD equations a valuable tool for
studying fundamental properties of MHD turbulence.  For instance,
simulations indicate that the scaling of the energy spectrum of strong
MHD turbulence is anisotropic, as in GS picture, but with the scaling
of the IK
phenomenology~\cite{maron,muller2,mason,mason2,perez-boldyrev}.  In
order to resolve this controversy, it has been
proposed~\cite{boldyrev,boldyrev2} that magnetic and velocity
fluctuations tend to align their polarizations in turbulent eddies,
leading to scale-dependent depletion of nonlinear interaction of
Alfv\'en modes. This leads to the anisotropic energy spectrum
$E(k_\perp)\sim k_\perp^{-3/2}$, in agreement with numerical
simulations.

The phenomenon of scale-dependent dynamic alignment is closely related 
to the conservation of cross-helicity, an ideal invariant 
cascading toward small scales in a
turbulent state. Cross-helicity has only recently become an object of systematic
study, as it become clear that it plays a fundamental role in driven
MHD turbulence~\citep{boldyrev,boldyrev2,mason,mason2,lithwick03,lithwick07,chandran08,beresnyak08,matthaeus08}. 

Denote $E^\pm=\langle |\vec z^\pm|^2\rangle /4$ the energy associated
with the $\pm$ waves, then the total energy and cross-helicity of the
system are $E=E^++E^-$ and $H_c=E^+-E^-$, respectively. Both energy
and cross-helicity are invariants of the ideal MHD
equations. Cross-helicity provides a measure of the imbalance between
non-linearly interacting waves. When it does not vanish the turbulence
is called \emph{imbalanced}, otherwise it is
\emph{balanced}. 

A significant interest to imbalanced MHD turbulence has also been
motivated by astrophysical solar wind data, which indicate that solar
wind turbulence is dominated by Alfv\'en waves propagating outward from the
sun~\citep[e.g.,][]{goldstein}. A number of analytic derivations of
the spectra of imbalanced MHD turbulence have been recently
proposed~\citep{lithwick03,lithwick07,beresnyak08,chandran08}, which
however lead to contradictory results. Most of numerical simulations
of MHD turbulence have so far concentrated on balanced cases, and
practically no systematic study of strong imbalanced MHD turbulence in
high-resolution direct numerical simulations has been available.

 

In an attempt to address the issue and to resolve the contradictions, 
we performed high resolution numerical simulations to investigate the
inertial range of strong MHD 
turbulence with and without cross-helicity.  Based on our results, we
propose that in the imbalanced case the Els\"asser energy spectra have
different amplitudes, nevertheless, their scaling is the same,
$E^+(k_{\perp})\propto E^-(k_{\perp})\propto k_{\perp}^{-3/2}$. This
scaling coincides with the scaling in the balanced case, which is
consistent with the view that balanced MHD turbulence is as a
superposition of locally imbalanced regions. We demonstrate that this
picture is essentially consistent with the phenomenon of
scale-dependent dynamic alignment, and provide an analytic explanation
for the obtained spectra. 
\vskip2mm

\emph{Model equations.}---
The universal properties of
weak and strong turbulence in MHD are accurately described by
neglecting the parallel component of the fluctuating fields,
associated with the pseudo-Alfv\'en
mode~\cite{galtier-chandran,perez-boldyrev}. 
By setting
$\vec z_\|^\pm = 0$ in equation \eref{mhd-elsasser} we obtain the
closed system of equations
\begin{eqnarray}
  \(\frac{\partial}{\partial t}\mp\vec v_A\cdot\nabla_\|\)\vec
  z^\pm+\left(\vec z^\mp\cdot\nabla_\perp\right)\vec z^\pm =
  -\nabla_\perp P\nonumber\\ +\vec f^\pm+\nu\nabla^2\vec z^\pm,
  \label{rmhd-elsasser}
\end{eqnarray}
in which force and dissipation terms have been added to address the
case of steadily driven turbulence, and we assume that viscosity is
equal to resistivity.  This set of equations is known as the Reduced
MHD model (RMHD) 
\cite{kadomtsev,strauss}. It is worth mentioning that the RMHD model
was derived as an approximation of the full MHD equations
in the limit $k_\| \ll k_\perp$, and therefore it is applicable to
strong turbulence. It has been recently realized that when used in a
broader $k_\|-k_{\perp}$ domain the system~(\ref{rmhd-elsasser})
describes the universal regime of {\em weak} Alfv\'enic
turbulence~\citep{perez-boldyrev}. This opens the possibility of
effective analytic and numerical study of both weak and strong MHD
turbulence in the same framework. In the present paper, we use this
system to study strong anisotropic MHD turbulence. 
\vskip2mm

\emph{Numerical method.}---We employ a fully dealiased Fourier
pseudo-spectral method to solve equations \eref{rmhd-elsasser} with a
strong guide field ($ v_A/v_{rms}\sim 5$) in a rectangular periodic
box, with field-perpendicular cross section $L_\perp^2=(2\pi)^2$ and
field-parallel box size $L_\|$. The choice of a rectangular box, as
discussed in~\cite{perez-boldyrev}, allows for the excitation of
elongated modes at large scales, necessary to avoid a long transition
region from the forcing to the inertial interval, which can lead to
inaccurate measurements of the spectral index.

In order to achieve a steady state, the random forcing $\vec f^\pm$ is
applied in Fourier space at wave-numbers $1 \leq k_{\perp} \leq 2, (2\pi/L_\|) 
\leq k_\|\leq (2\pi/L_\|)n_\|$, where $n_\|$ controls the width of the
force spectrum in $k_\|$. The Fourier coefficients inside that range
are Gaussian random numbers with amplitude chosen so that the
resulting rms velocity fluctuations are of order unity.  The
individual random values are refreshed independently for each mode on
average every $\tau=0.1~L_\perp/v_{rms}$. As shown
in~\cite{perez-boldyrev}, the width of the field-parallel spectrum
controls the critical balance at the forcing scale, and determines
whether the turbulence is weak or strong. We define the Reynolds
number as $Re=(L_\perp/2\pi)v_{rms}/\nu$.

In the present simulations, we also introduce correlation between
$\vec v$ and $\vec b$, to investigate the role of cross-helicity. Such
correlation is introduced at the forcing scales by controlling the
correlation between the velocity and magnetic field forces, $\vec f_v$
and $\vec f_b$. This is achieved by taking $\vec f^\pm$ as
uncorrelated Gaussian random forces, so that $ \vec f_v = \frac
12\(\vec f^++\vec f^-\)$, $\vec f_b = \frac 12\(\vec f^+-\vec f^-\)$,
from which it immediately follows that cross-helicity is controlled by
the difference in the variances: $ \langle\vec
f_v\cdot\vec f_b \rangle = \frac 14\(\sigma_+^2-\sigma_-^2\)$, where
$\sigma^2_\pm\equiv \langle |{\bf f}^\pm|^2\rangle$.
\begin{table}[!tb]
\begin{tabular}{c@{\hspace{0.5cm}}c@{\hspace{0.5cm}}c@{\hspace{0.5cm}}c@{\hspace{0.5cm}}c@{\hspace{0.5cm}}c@{\hspace{0.5cm}\  
\ }c@{\hspace{0.5cm}}} \hline\hline Run & Resolution & $Re$ & $\alpha$
& $L_\|/L_\perp$ & $n_\|$ & $\rho_c$ \\ \hline A & $1024^2\times 256$ &
4000 & 0.  & 6 & 1 & 0 \\ B & $512^2\times 256$ & 1500 & 0.3 & 5 & 2 &
0.6 \\ C1 & $512^3$ & 1500 & 0.25 & 10 & 2 & 0.6 \\ C2 & $1024^2\times256$ & 4000 & 0.25 & 10 & 2 & 0.6\\ \hline\hline
  \end{tabular}
  \caption{Summary of simulations. A, B, C correspond to strong
  turbulence with different amount of net cross-helicity.}
  \label{sims_table}
\end{table}
It is convenient to define the parameter $\alpha$ and the normalized
cross helicity:
\begin{eqnarray}
\alpha&\equiv& (\sigma_+^2-\sigma_-^2)/(\sigma_+^2+\sigma_-^2)=
2\langle\vec f_v\cdot\vec f_b\rangle,\\
\rho_c&\equiv& H_c/E=(E^+-E^-)/(E^++E^-)
\end{eqnarray}
\vskip2mm

\begin{figure}
  \includegraphics[width=0.45\textwidth]{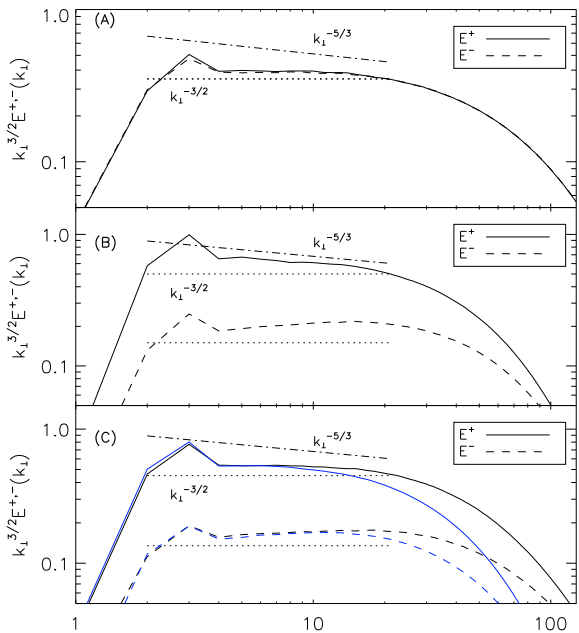}
\vskip-10mm
  \caption{Spectra of balanced and imbalanced strong turbulence corresponding to runs A, B, C1 (blue) and C2 (black) (see the text). 
  In plots (B) and (C), the spectrum of $z^{-}$ is arbitrarily offset in the vertical direction.}
  \label{strong}
\end{figure}

\emph{Results.}---Table \ref{sims_table} shows the summary of three
representative simulations. Runs A, B, C were carried out with narrow
$k_\|$-band forcing that produces critically balanced and strongly
interacting large-scale modes.  The energy spectra are shown in
Fig.~\ref{strong}, compensated by $k_\perp^{3/2}$.  
In run A with $\alpha=0$, we observe a 
balanced turbulence with the scaling close to $E^+\sim E^-\sim
k_\perp^{-3/2}$. When cross-helicity is introduced in Run B, we
observe a slight steepening of $E^+\sim k_\perp^{-1.6}$ and a slight
flattening of $E^-\sim k_\perp^{-1.35}$. This behavior is justified as
follows: since the total energy $E=E^++E^-$ is kept constant, when the
cross-helicity increases, the amplitude of $\vec z^+$ increases at the
expense of $\vec z^-$.  Therefore, the nonlinear interaction of $\vec
z^+$ with $\vec z^-$ becomes weaker, resulting in a steepening of the 
spectrum. This steepening is however an artifact of a not optimal 
numerical setting. To simulate this interaction correctly, we need to
elongate the box in field-parallel direction so as to fit the eddies
with longer parallel wavelengths at the forcing scales. As a result,
the $E^\pm$ spectra get closer to $k_\perp^{-3/2}$; this is evident in
Runs C1 and C2.  Note that the limit of very large cross-helicity would require
extremely long simulation box in order to observe the universal
scaling behavior $k_\perp^{-3/2}$.
\vskip2mm

\begin{figure}
  \begin{center}
  \includegraphics[width=3.5in]{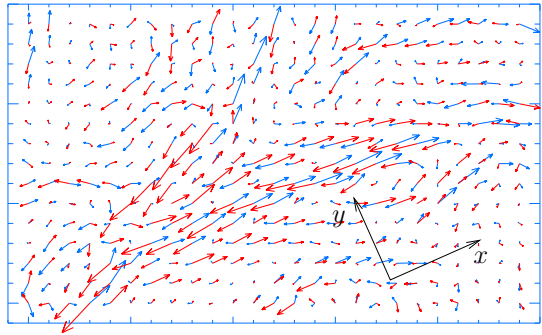}
  \end{center}
  \vspace*{-0.5cm}
  \caption{A correlated region of (counter-)aligned magnetic and
velocity fluctuations (red and blue vectors) at scale $\lambda=L_\perp/12$, in a
plane perpendicular to the strong guide field, in run A. The fluctuations are aligned 
predominantly in the $x$~direction while their directions and amplitudes change 
predominantly in the~$y$ direction.}\label{domains_corr}
\end{figure}

\emph{Discussion.}---In this section we propose an explanation for the
observed spectra. Our explanation essentially relies on the phenomenon
of scale-dependent dynamic alignment. To understand how the alignment
affects the energy spectrum, consider the eddy shown in
Fig.~\ref{domains_corr}. In this eddy fluctuations are aligned 
within the small angle $\theta_{\lambda}$, 
while their directions and
magnitudes change in an almost perpendicular direction.  In the case
of strong balanced turbulence, the nonlinear interaction in such an
eddy is then reduced by a factor $\theta_{\lambda}$ for both $z^+$ and
$z^-$ fields, and the corresponding nonlinear interaction time is
estimated as $\tau_{\lambda} \sim 1/({\bf z}^{\pm}_{\lambda} \cdot
{\bf k}_{\perp}) \sim 1/(z^{\pm}_{\lambda}k_{\perp}
\theta_{\lambda})$. The scaling of the fluctuating fields is then
found from the requirement of constant energy fluxes:
$(z_{\lambda}^{\pm})^2/\tau_{\lambda}={\rm const}$.  One can argue
\citep{boldyrev2,mason,mason2} that the alignment angle decreases with
scale as $\theta_{\lambda}\propto \lambda^{1/4}$, in which case the
field-perpendicular energy spectrum is $E(k_{\perp})\propto
k_{\perp}^{-3/2}$.
\begin{figure}
	\centering \includegraphics[width=1.8cm,
		angle=-90]{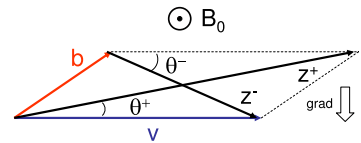}
	\caption{Sketch of dynamic alignment of magnetic and velocity
	fluctuations in a turbulent eddy.}
	\label{fig:angles}
\end{figure}

In the case of strong imbalanced turbulence, the alignment is still
preserved. However, since the fields amplitudes are essentially
different the alignment angles are different as well; we denote them
$\theta_{\lambda}^+$ and $\theta_{\lambda}^-$, see Fig.~\ref{fig:angles}. 
The assumption of the
dynamic alignment then leads to the important geometric constraint:
$\theta_{\lambda}^+z_{\lambda}^+\sim \theta_{\lambda}^-z_{\lambda}^-$,
as is clear from Fig.~\ref{fig:angles}.  The depletion of nonlinear
interaction is therefore different for $z^+$ and $z^-$ fields,
however, their nonlinear interaction times, $\tau^{\mp}_{\lambda} \sim
1/(z^{\pm}_{\lambda}k_{\perp} \theta^{\pm}_{\lambda})$, are the
same. The requirement of constant energy fluxes
$(z_{\lambda}^{\pm})^2/\tau^{\pm}_{\lambda}\sim \epsilon^{\pm}={\rm
const}$ then ensures that $z_{\lambda}^+/z_{\lambda}^-\sim
\sqrt{\epsilon^+/\epsilon^-}$, so both fields should have the same
scaling, although different amplitudes. The geometric constraint then
leads to $\theta_{\lambda}^+/\theta_{\lambda}^-\sim
\sqrt{\epsilon^-/\epsilon^+}$, so the alignment angles should have the
same scaling as well.


To conclude this section we compare our results with recent analytic
predictions of~\citet{lithwick07}, \citet{beresnyak08}, and
\citet{chandran08}. 
The main difference of our model with previous studies is that we 
include the phenomenon of dynamic alignment. In~\citep{lithwick07} the
energy cascade times were assumed 
to be essentially different and the derived spectra had the form
$E^+(k_{\perp})\propto E_{\perp}^-\propto k_{\perp}^{-5/3}$,  while
our numerical results are more consistent with
$k_\perp^{-3/2}$. In~\citep{beresnyak08,chandran08}, it was  
assumed that the $z^+$ field undergoes a weak cascade, while $z^-$ a
strong cascade,  
leading to different spectra of~$z^+$
and~$z^-$, which seems to be supported by numerical simulations in 
~\citep{beresnyak08}. We however note that the steepening of the $z^+$
and the flattening of the $z^-$ spectra  
might be due to the high level of cross-helicity in these simulations (around
$\rho_c\sim 0.97$), which requires an extremely elongated simulation box
in the field-parallel direction, cf. our Fig.~\ref{strong}(B). In
addition, the simulations in~\citep{beresnyak08} are 
performed on lower resolution with hyperviscosity,
which might alter the spectra due to a bottleneck effect.

\begin{figure}[!htb]
  \includegraphics[width=3.3in]{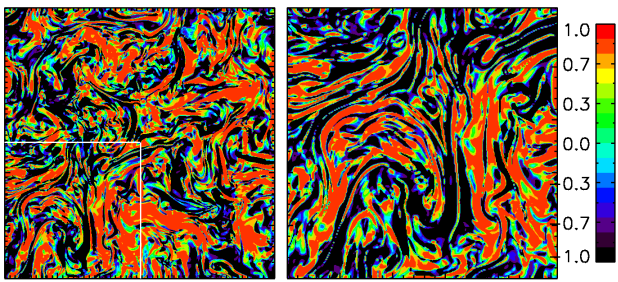}
  \caption{Cosine of the alignment angle between ${\bf v}_{\lambda}$
  and ${\bf b}_{\lambda}$ fluctuations in the guide-field
  perpendicular plane at scales $\lambda=L_\perp/6$ (left), and
  $\lambda=L_\perp/12$ (right) in Run A. The right frame corresponds
  to the region delimited by the white square on the left side.}
  \label{domains}
\end{figure}

{\em Conclusion.}---We have presented the results of numerical
simulations of strong MHD turbulence in both balanced and imbalanced
regimes.  In the imbalanced turbulence, say with positive
cross-helicity, the total energy spectrum $E=E^++E^-$ is dominated by
$E^+$. Simulations in this case show a universal inertial-range
regime: although the $E^+$ and $E^-$ spectra have different
amplitudes, their scaling is the same $E^+(k_{\perp})\propto
E^-(k_{\perp})\propto k_{\perp}^{-3/2}$.

In the balanced turbulence, both spectra have the same amplitudes and
scaling $E^+(k_\perp)\sim E^-(k_\perp)\sim k_\perp^{-3/2}$. This is
consistent with the view that overall balanced turbulence can be
imbalanced locally, creating patches (eddies) of positive and negative
cross-helicity. In each of these regions the picture of imbalanced
turbulence presented above applies. When averaged over all the
regions, the spectra of balanced turbulence are reproduced.

The presented picture of MHD turbulence is consistent with the
phenomenon of scale-dependent dynamic
alignment~\cite{boldyrev,boldyrev2}, which provides a natural
explanation for the observed spectra. In this phenomenology, the
configuration space splits into eddies with highly aligned and
anti-aligned magnetic and velocity fluctuations, where nonlinear
interactions are reduced. Fig.~\ref{domains} shows the cosine of the
alignment angle between velocity and magnetic fluctuations for the
balanced simulations (run A), at scales $\lambda=L_\perp/6$ and
$\lambda=L_\perp/12$. The alignment 
possesses a hierarchical structure: inside small eddies
there exist smaller and more anisotropic eddies (of both
polarities), and so on. This hierarchical structure, first observed
by Mason \& Cattaneo [unpublished, 2006], presents an interesting example 
of magnetic self-organization in a driven turbulent system.

\acknowledgments This work was supported by the U.S.  DOE under Grant
No.~DE-FG02-07ER54932, by the NSF Center for 
Magnetic Self-Organization in Laboratory and Astrophysical Plasmas at
the UW-Madison, and in part by the NSF under
Grant No. NSF PHY05-51164. High Performance Computing resources were
provided by the Texas Advanced Computing Center (TACC) at the
University of Texas at Austin under the NSF-Teragrid Project
TG-PHY080013N.

\bibliographystyle{apsrev}

\begin{thebibliography}{27}
\expandafter\ifx\csname natexlab\endcsname\relax\def\natexlab#1{#1}\fi
\expandafter\ifx\csname bibnamefont\endcsname\relax
  \def\bibnamefont#1{#1}\fi
\expandafter\ifx\csname bibfnamefont\endcsname\relax
  \def\bibfnamefont#1{#1}\fi
\expandafter\ifx\csname citenamefont\endcsname\relax
  \def\citenamefont#1{#1}\fi
\expandafter\ifx\csname url\endcsname\relax
  \def\url#1{\texttt{#1}}\fi
\expandafter\ifx\csname urlprefix\endcsname\relax\def\urlprefix{URL }\fi
\providecommand{\bibinfo}[2]{#2}
\providecommand{\eprint}[2][]{\url{#2}}

\bibitem[{\citenamefont{Biskamp}(2003)}]{biskamp}
\bibinfo{author}{\bibfnamefont{D.}~\bibnamefont{Biskamp}},
  \emph{\bibinfo{title}{Magnetohydrodynamic Turbulence}}
  (\bibinfo{publisher}{Cambridge University Press, Cambridge},
  \bibinfo{year}{2003}).

\bibitem[{\citenamefont{Iroshnikov}(1963)}]{iroshnikov63}
\bibinfo{author}{\bibfnamefont{P.~S.} \bibnamefont{Iroshnikov}},
  \bibinfo{journal}{AZh} \textbf{\bibinfo{volume}{40}}, \bibinfo{pages}{742}
  (\bibinfo{year}{1963}).

\bibitem[{\citenamefont{Iroshnikov}(1964)}]{iroshnikov64}
\bibinfo{author}{\bibfnamefont{P.~S.} \bibnamefont{Iroshnikov}},
  \bibinfo{journal}{Sov. Astron} \textbf{\bibinfo{volume}{7}},
  \bibinfo{pages}{566} (\bibinfo{year}{1964}).

\bibitem[{\citenamefont{Kraichnan}(1965)}]{kraichnan}
\bibinfo{author}{\bibfnamefont{R.~H.} \bibnamefont{Kraichnan}},
  \bibinfo{journal}{Phys. Fluids} \textbf{\bibinfo{volume}{8}},
  \bibinfo{pages}{1385} (\bibinfo{year}{1965}).

\bibitem[{\citenamefont{Shebalin et~al.}(1983)\citenamefont{Shebalin,
  Mattheaus, and Montgomery}}]{Shebalin}
\bibinfo{author}{\bibfnamefont{J.~V.} \bibnamefont{Shebalin}},
  \bibinfo{author}{\bibfnamefont{W.~H.} \bibnamefont{Mattheaus}},
  \bibnamefont{and} \bibinfo{author}{\bibfnamefont{D.~J.}
  \bibnamefont{Montgomery}}, \bibinfo{journal}{J.~Plasma Physics}
  \textbf{\bibinfo{volume}{29}}, \bibinfo{pages}{525} (\bibinfo{year}{1983}).

\bibitem[{\citenamefont{Montgomery and Turner}(1981)}]{Montgomery81}
\bibinfo{author}{\bibfnamefont{D.}~\bibnamefont{Montgomery}} \bibnamefont{and}
  \bibinfo{author}{\bibfnamefont{L.}~\bibnamefont{Turner}},
  \bibinfo{journal}{Phys. Fluids} \textbf{\bibinfo{volume}{24}},
  \bibinfo{pages}{825} (\bibinfo{year}{1981}).

\bibitem[{\citenamefont{Galtier et~al.}(2000)\citenamefont{Galtier, Nazarenko,
  Newell, and Pouquet}}]{galtier}
\bibinfo{author}{\bibfnamefont{S.}~\bibnamefont{Galtier}},
  \bibinfo{author}{\bibfnamefont{S.~V.} \bibnamefont{Nazarenko}},
  \bibinfo{author}{\bibfnamefont{A.~C.} \bibnamefont{Newell}},
  \bibnamefont{and} \bibinfo{author}{\bibfnamefont{A.}~\bibnamefont{Pouquet}},
  \bibinfo{journal}{J.~Plasma Physics} \textbf{\bibinfo{volume}{63}},
  \bibinfo{pages}{447} (\bibinfo{year}{2000}).

\bibitem[{\citenamefont{Ng and Bhattacharjee}(1996)}]{ng-bhattacharjee}
\bibinfo{author}{\bibfnamefont{C.~S.} \bibnamefont{Ng}} \bibnamefont{and}
  \bibinfo{author}{\bibfnamefont{A.}~\bibnamefont{Bhattacharjee}},
  \bibinfo{journal}{ApJ} \textbf{\bibinfo{volume}{465}}, \bibinfo{pages}{845}
  (\bibinfo{year}{1996}).

\bibitem[{\citenamefont{Bhattacharjee and Ng}(2001)}]{bhattacharjee-ng}
\bibinfo{author}{\bibfnamefont{A.}~\bibnamefont{Bhattacharjee}}
  \bibnamefont{and} \bibinfo{author}{\bibfnamefont{C.~S.} \bibnamefont{Ng}},
  \bibinfo{journal}{ApJ} \textbf{\bibinfo{volume}{548}}, \bibinfo{pages}{318}
  (\bibinfo{year}{2001}).

\bibitem[{\citenamefont{Ng et~al.}(2003)\citenamefont{Ng, Bhattacharjee,
  Germashewski, and Galtier}}]{ng}
\bibinfo{author}{\bibfnamefont{C.~S.} \bibnamefont{Ng}},
  \bibinfo{author}{\bibfnamefont{A.}~\bibnamefont{Bhattacharjee}},
  \bibinfo{author}{\bibfnamefont{K.}~\bibnamefont{Germashewski}},
  \bibnamefont{and} \bibinfo{author}{\bibfnamefont{S.}~\bibnamefont{Galtier}},
  \bibinfo{journal}{Phys. Plasmas} \textbf{\bibinfo{volume}{10}},
  \bibinfo{pages}{1954} (\bibinfo{year}{2003}).

\bibitem[{\citenamefont{Goldreich and Sridhar}(1995)}]{GS95}
\bibinfo{author}{\bibfnamefont{P.}~\bibnamefont{Goldreich}} \bibnamefont{and}
  \bibinfo{author}{\bibfnamefont{S.}~\bibnamefont{Sridhar}},
  \bibinfo{journal}{ApJ} \textbf{\bibinfo{volume}{438}}, \bibinfo{pages}{763}
  (\bibinfo{year}{1995}).

\bibitem[{\citenamefont{Maron and Goldreich}(2001)}]{maron}
\bibinfo{author}{\bibfnamefont{J.}~\bibnamefont{Maron}} \bibnamefont{and}
  \bibinfo{author}{\bibfnamefont{P.}~\bibnamefont{Goldreich}},
  \bibinfo{journal}{ApJ} \textbf{\bibinfo{volume}{554}}, \bibinfo{pages}{1175}
  (\bibinfo{year}{2001}).

\bibitem[{\citenamefont{M\"uller and Grappin}(2005)}]{muller2}
\bibinfo{author}{\bibfnamefont{W.-C.} \bibnamefont{M\"uller}} \bibnamefont{and}
  \bibinfo{author}{\bibfnamefont{R.}~\bibnamefont{Grappin}},
  \bibinfo{journal}{Phys. Rev. Lett.} \textbf{\bibinfo{volume}{95}},
  \bibinfo{pages}{114502} (\bibinfo{year}{2005}).

\bibitem[{\citenamefont{Mason et~al.}(2006)\citenamefont{Mason, Cattaneo, and
  Boldyrev}}]{mason}
\bibinfo{author}{\bibfnamefont{J.}~\bibnamefont{Mason}},
  \bibinfo{author}{\bibfnamefont{F.}~\bibnamefont{Cattaneo}}, \bibnamefont{and}
  \bibinfo{author}{\bibfnamefont{S.}~\bibnamefont{Boldyrev}},
  \bibinfo{journal}{Phys. Rev. Lett.} \textbf{\bibinfo{volume}{97}},
  \bibinfo{pages}{255002} (\bibinfo{year}{2006}).

\bibitem[{\citenamefont{Mason et~al.}(2007)\citenamefont{Mason, Cattaneo, and
  Boldyrev}}]{mason2}
\bibinfo{author}{\bibfnamefont{J.}~\bibnamefont{Mason}},
  \bibinfo{author}{\bibfnamefont{F.}~\bibnamefont{Cattaneo}}, \bibnamefont{and}
  \bibinfo{author}{\bibfnamefont{S.}~\bibnamefont{Boldyrev}},
  \bibinfo{journal}{arXiv:0706.2003}  (\bibinfo{year}{2007}).

\bibitem[{\citenamefont{Perez and Boldyrev}(2008)}]{perez-boldyrev}
\bibinfo{author}{\bibfnamefont{J.~C.} \bibnamefont{Perez}} \bibnamefont{and}
  \bibinfo{author}{\bibfnamefont{S.}~\bibnamefont{Boldyrev}},
  \bibinfo{journal}{ApJ} \textbf{\bibinfo{volume}{672}}, \bibinfo{pages}{L61}
  (\bibinfo{year}{2008}).

\bibitem[{\citenamefont{Boldyrev}(2005)}]{boldyrev}
\bibinfo{author}{\bibfnamefont{S.}~\bibnamefont{Boldyrev}},
  \bibinfo{journal}{ApJ} \textbf{\bibinfo{volume}{626}}, \bibinfo{pages}{L37}
  (\bibinfo{year}{2005}).

\bibitem[{\citenamefont{Boldyrev}(2006)}]{boldyrev2}
\bibinfo{author}{\bibfnamefont{S.}~\bibnamefont{Boldyrev}},
  \bibinfo{journal}{Phys. Rev. Lett.} \textbf{\bibinfo{volume}{96}},
  \bibinfo{pages}{115002} (\bibinfo{year}{2006}).

\bibitem[{\citenamefont{Lithwick and Goldreich}(2003)}]{lithwick03}
\bibinfo{author}{\bibfnamefont{Y.}~\bibnamefont{Lithwick}} \bibnamefont{and}
  \bibinfo{author}{\bibfnamefont{P.}~\bibnamefont{Goldreich}},
  \bibinfo{journal}{ApJ} \textbf{\bibinfo{volume}{582}}, \bibinfo{pages}{1220}
  (\bibinfo{year}{2003}).

\bibitem[{\citenamefont{Lithwick et~al.}(2007)\citenamefont{Lithwick,
  Goldreich, and Sridhar}}]{lithwick07}
\bibinfo{author}{\bibfnamefont{Y.}~\bibnamefont{Lithwick}},
  \bibinfo{author}{\bibfnamefont{P.}~\bibnamefont{Goldreich}},
  \bibnamefont{and} \bibinfo{author}{\bibfnamefont{S.}~\bibnamefont{Sridhar}},
  \bibinfo{journal}{ApJ} \textbf{\bibinfo{volume}{655}}, \bibinfo{pages}{269}
  (\bibinfo{year}{2007}).

\bibitem[{\citenamefont{Chandran}(2008)}]{chandran08}
\bibinfo{author}{\bibfnamefont{B.}~\bibnamefont{Chandran}},
  \bibinfo{journal}{Astrophys. J.} \textbf{\bibinfo{volume}{685}}, \bibinfo{pages}{646}   (\bibinfo{year}{2008}).

\bibitem[{\citenamefont{Beresnyak and Lazarian}(2008)}]{beresnyak08}
\bibinfo{author}{\bibfnamefont{A.}~\bibnamefont{Beresnyak}} \bibnamefont{and}
  \bibinfo{author}{\bibfnamefont{A.}~\bibnamefont{Lazarian}},
  \bibinfo{journal}{ApJ} \textbf{\bibinfo{volume}{682}}, \bibinfo{pages}{1070}
  (\bibinfo{year}{2008}).

\bibitem[{\citenamefont{Matthaeus et~al.}(2008)\citenamefont{Matthaeus,
  Pouquet, Mininni, Dmitruk, and Breech}}]{matthaeus08}
\bibinfo{author}{\bibfnamefont{W.~H.} \bibnamefont{Matthaeus}},
  \bibinfo{author}{\bibfnamefont{A.}~\bibnamefont{Pouquet}},
  \bibinfo{author}{\bibfnamefont{P.~D.} \bibnamefont{Mininni}},
  \bibinfo{author}{\bibfnamefont{P.}~\bibnamefont{Dmitruk}}, \bibnamefont{and}
  \bibinfo{author}{\bibfnamefont{B.}~\bibnamefont{Breech}},
  \bibinfo{journal}{Phys. Rev. Lett.} \textbf{\bibinfo{volume}{100}},
  \bibinfo{pages}{085003} (\bibinfo{year}{2008}).

\bibitem[{\citenamefont{Goldstein et~al.}(1995)\citenamefont{Goldstein,
  Roberts, and Matthaeus}}]{goldstein}
\bibinfo{author}{\bibfnamefont{D.~A.} \bibnamefont{Goldstein}},
  \bibinfo{author}{\bibfnamefont{D.~A.} \bibnamefont{Roberts}},
  \bibnamefont{and} \bibinfo{author}{\bibfnamefont{W.~H.}
  \bibnamefont{Matthaeus}}, \bibinfo{journal}{Annu. Rev. Astron. Astrophys.}
  \textbf{\bibinfo{volume}{33}}, \bibinfo{pages}{283} (\bibinfo{year}{1995}).

\bibitem[{\citenamefont{Galtier and Chandran}(2006)}]{galtier-chandran}
\bibinfo{author}{\bibfnamefont{S.}~\bibnamefont{Galtier}} \bibnamefont{and}
  \bibinfo{author}{\bibfnamefont{B.~D.~G.} \bibnamefont{Chandran}},
  \bibinfo{journal}{Phys. Plasmas} \textbf{\bibinfo{volume}{13}},
  \bibinfo{pages}{114505} (\bibinfo{year}{2006}).

\bibitem[{\citenamefont{Kadomtsev and Pogutse}(1974)}]{kadomtsev}
\bibinfo{author}{\bibfnamefont{B.~B.} \bibnamefont{Kadomtsev}}
  \bibnamefont{and} \bibinfo{author}{\bibfnamefont{O.~P.}
  \bibnamefont{Pogutse}}, \bibinfo{journal}{Sov. Phys.- JETP}
  \textbf{\bibinfo{volume}{38}}, \bibinfo{pages}{283} (\bibinfo{year}{1974}).

\bibitem[{\citenamefont{Strauss}(1976)}]{strauss}
\bibinfo{author}{\bibfnamefont{H.~R.} \bibnamefont{Strauss}},
  \bibinfo{journal}{Phys. Fluids} \textbf{\bibinfo{volume}{19}},
  \bibinfo{pages}{134} (\bibinfo{year}{1976}).

\end{thebibliography}

\end {document}